\documentstyle[preprint,aps,]{revtex}

\begin{document}
\def\bea{\begin{eqnarray}} 
\def\eea{\end{eqnarray}} 
\def\ap{\approx}
\def\barr#1{\begin{array}{#1}} 
\def\earr{\end{array}} 
\def\l{\lambda} 
\def\L{\Lambda}
\def\hl{\hat{\lambda}} 
\def\hL{\hat{\Lambda}}
\def\gtr{\gtrsim}
\def\lsr{\lesssim}
\def\g{(g^2+g^{\prime 2})}
\def\g1{{\tilde{g}}^2}
\def\la{\langle}
\def\ra{\rangle}
\def\a{\alpha}
\def\p{\pi}
\def\pr{\prime}
\def\f{\frac}
\def\m#1{m_{\rm #1}^2}
\def\nn{\nonumber} 
\tightenlines

\preprint{\begin{tabular}{r} KAIST-TH 99/04 \\ hep-ph/9906363 \end{tabular}}

\title{
Dynamical solution to the $\mu$ problem at TeV scale
}

\author{
Kiwoon Choi\thanks{kchoi@higgs.kaist.ac.kr}
and Hyung Do Kim\thanks{hdkim@muon.kaist.ac.kr}
}

\address{
Department of Physics, Korea Advanced Institute of Science and Technology \\
Taejon 305-701, Korea \\
}

\maketitle

\begin{abstract}
We introduce a new confining force ($\mu$-color) at TeV scale 
to dynamically generate a supersymmetry preserving mass scale
which would replace the $\mu$ parameter in the minimal supersymmetric 
standard model (MSSM). 
We discuss the Higgs phenomenology and also
the pattern of soft supersymmetry breaking parameters 
allowing   the correct electroweak symmetry breaking within
the $\mu$-color model, which have quite distinctive features 
from the MSSM and also from other generalizations of the MSSM.
\end{abstract}

\pacs{PACS Number(s): }

\section{Introduction}
The minimal supersymmetric standard model (MSSM)
contains  two different  types of mass scales:
(i) soft supersymmetry (SUSY) breaking parameters $m_{\rm soft}$ 
including  the soft scalar and gaugino masses
and (ii) the $\mu$ parameter in the superpotential
$W \ni \mu H_1 H_2$ where $H_1$ and $H_2$ are the MSSM Higgs
doublets with opposite hypercharge.
In the MSSM point of view, 
$\mu$ is entirely different from
$m_{\rm soft}$ since it has nothing to do with
SUSY breaking\cite{kim-nilles}.
In order to have  correct electroweak symmetry breaking without severe
fine tuning,
both $m_{\rm soft}$ and $\mu$ are required to
be of order the electroweak scale.
Although it is technically natural that both $m_{\rm soft}$
and $\mu$ are much smaller than the cutoff scale of the model
which may be as large as
the Planck scale $M_{Pl}$, one still needs  
to understand the dynamical origin of these mass scales
for deeper understanding of their smallness\cite{nilles,kim-nilles}.

It is commonly assumed that $m_{\rm soft}$ arises as a consequence
of spontaneous SUSY breaking at high energy scales.
The explicit relation between $m_{\rm soft}$
and the scale of spontaneous SUSY breaking
depends on how the SUSY breaking is transmitted to the MSSM sector:
(i) $m_{\rm soft}\sim F/M_{Pl}$ in the case of gravity mediation
with SUSY breaking auxiliary component $F$\cite{nilles}
and (ii) $m_{\rm soft}\sim (\frac{\alpha}{4\pi}) F/M_X$ in the case of
gauge mediation\cite{dine} by a messenger particle with mass $M_X$.
In both cases  $\sqrt{F}$ is significantly larger than 1 TeV,
$\sqrt{F}\sim 10^{8}$ TeV for gravity-mediated case 
and $\sqrt{F}\gtr 20$ TeV
for gauge-mediated case\cite{GMM},
so it is quite unlikely that SUSY breaking dynamics can be directly probed 
by future experiments. 

About the dynamical origin of $\mu$, there 
have been   many interesting suggestions in the literatures
\cite{mu1,mu2,mu3,mu4,mu5,murayama}.
Perhaps the most attractive possibility
would be that SUSY breaking dynamics provides a dynamical
seed for both $\mu$ and $m_{\rm soft}$ in a manner
to yield $\mu\sim m_{\rm soft}$.
In most cases, these schemes are again  based on high energy dynamics
which is hard to be probed by future experiments.
In this paper we wish to propose an alternative scheme  
replacing $\mu$  by a new confining force 
($\mu$-color) at TeV scale  which would lead to interesting
phenomenologies in future experiments. 

The $\mu$ term  is essential in the MSSM for several 
phenomenological reasons.
Its absence implies the absence of the 
associated $B$-term ($B\mu H_1H_2$) in the scalar potential,
leading to  $\langle H_1\rangle=0$ even 
when  nonzero  $\langle H_2\rangle$ is radiatively induced
by the large top quark Yukawa coupling
and also to the  phenomenologically 
unacceptable Weinberg-Wilczek axion\cite{kim-nilles}.
The $\mu$ term is necessary  also
to render sufficiently large masses to the Higgsinos.
In the $\mu$-color model, Yukawa couplings of
$H_{1,2}$ with the $\mu$-colored matter fields generate
{\it effective} $\mu$ terms involving
the composite Higgs doublets.
The unwanted axion is avoided due to the $U(1)_{PQ}$ breaking
by the strong $\mu$-color anomaly,
and also the correct electroweak symmetry breaking
can be achieved by the combined effects of the $\mu$-color dynamics
and  soft SUSY breaking terms.

As we will see, the $\mu$-color model is distinguished
from the MSSM (and also from many  other generalizations of the MSSM)
mainly by its Higgs sector.
It is distinguished also by the pattern of soft parameters
which would allow the correct electroweak symmetry breaking to take place.
Some soft parameter values which would lead to a successful
electroweak symmetry breaking within the MSSM can not
work within the $\mu$-color model,
while others which would not work within the MSSM
do work in the $\mu$-color model.
For instance, in the $\mu$-color case
it is not necessary to have a negative mass squared
of $H_1$ or $H_2$ for the electroweak symmetry breaking
to take place.
As another example of the difference,
a large portion of the  $(\tan\beta, M_{\rm m})$ space
in gauge-mediated SUSY breaking models appears to be 
incompatible with the $\mu$-color model where $M_{\rm m}$ is
the messenger scale of SUSY breaking, though it
can be compatible with the conventional $\mu$-term 
in the MSSM\cite{murayama}.
A potentially unattractive feature of the
$\mu$-color model is that it requires 
that the $\mu$-color gaugino mass at the messenger scale
is significantly smaller than the MSSM soft parameters
(by the factor of $1/16\pi^2$).
In gauge-mediated SUSY breaking models,
such a small $\mu$-color gaugino mass can  be achieved
if the messenger particles are $SU(2)_{\mu}$-singlets.
In gravity-mediated case, e.g. string effective supergravity
in which SUSY breaking is mediated by string moduli,
the $\mu$-color gaugino mass is small
if the $\mu$-color gauge kinetic function
does {\it not} depend on the messenger moduli 
at string tree level.
Thus the small $\mu$-color gaugino mass
may not be a serious drawback of the model.
At any rate, we note that string effective supergravity models
provide  large  varieties in the pattern of 
soft parameters\cite{soft1,soft2},
which are diverse enough to include those giving the correct 
Higgs phenomenology within the $\mu$-color model.

\section{the model}
The minimal $\mu$-color model includes, in addition to the MSSM 
gauge and matter multiplets,
the $\mu$-color gauge group  $SU(2)_{\mu}$
which confines at $\Lambda_{\mu}\sim 1$ TeV
and also the $\mu$-colored matter superfields which
transform under $SU(2)_{\mu}\times SU(2)_L\times U(1)_Y$ as
\bea
\label{assign}
Y_{\alpha a}=(2,2)_{0},
\quad
X_{1 a}=(2,1)_{1/2},
\quad
X_{2 a}=(2,1)_{-1/2},
\eea
where $a=1,2$ and $\alpha=1,2$ denote the $SU(2)_{\mu}$ and
$SU(2)_L$ doublet indices, respectively,
and the subscripts of the brackets  denote the $U(1)_Y$ charge.
Obviously these  additional  matters 
are free from  (both perturbative and global) 
gauge and gravitational anomalies.
The MSSM matter parity 
can be easily generalized to the $\mu$-color model  such that
the two MSSM Higgs doublets are even while all other matter multiplets
are odd under the generalized matter parity.
Then  the most general {\it scale-free} tree-level superpotential 
with the generalized matter parity  is given by
\bea
W_{\rm tree} & =& \l_1 H_1 Y X_1 + \l_2 H_2 Y X_2 \nn \\
 & & + \l_d H_1 Q D^c + \l_u H_2 Q U^c + \l_l H_1 L E^c,
\eea
where $H_{1,2}$, $Q$, $U^c$, $D^c$, $L$ and $E^c$
denote the MSSM fields in self-explanatory notation,
and all the gauge and  generation indices  are omitted here.

For the $\mu$-colored matter contents of Eq. (\ref{assign}),
the {\it holomorphic} $\mu$-color scale is given by \cite{seiberg}
\bea
\Lambda_{\mu}=M_{GUT}\exp (-\frac{2\pi^2}{g_{\mu}^2(M_{GUT})}+i
\frac{\theta_{\mu}}{4}),
\eea
where $g_{\mu}$ and $\theta_{\mu}$ are the $\mu$-color gauge coupling
and vacuum angle, respectively.
Once the extra matter multiplets of (\ref{assign}) carrying
$SU(2)_L\times U(1)_Y$ charges are introduced, we lose the
unification of gauge couplings at single energy scale.
However this may not be a serious drawback of the model
since there are many  string theory models, e.g. heterotic string theory
with a large threshold effects\cite{threshold} and/or Type I strings with
different type of D-branes\cite{munoz},
implying  that the gauge couplings at the string or unification scale
can  take different values.
At any rate, we note that  $\alpha_{\mu}(M_{GUT})\sim
1/19$ and $M_{GUT}\sim 10^{16}$ GeV lead to $\Lambda_{\mu}
\sim 1$ TeV, so having $\Lambda_{\mu}$ at TeV scale is
a plausible possibility.

A crucial feature of the $\mu$-color model is that
there is no mass parameter in $W_{\rm tree}$\footnote{
This may be explained by the $U(1)_R$ symmetry of Eq. (\ref{global}) 
which forbids the bilinear terms such as 
$YY$, $X_1X_2$, and $H_1H_2$ in the superpotential.}.
Thus at scales above $\Lambda_{\mu}$,
all the mass parameters are in
the soft SUSY breaking terms which are presumed
to be induced by  SUSY breaking  dynamics  at scales 
far above $\Lambda_{\mu}$.
For the scale-free tree level superpotential
$W_{\rm tree}=\lambda_{ijk}\Phi_i\Phi_j\Phi_k$, 
soft SUSY breaking terms can be written as 
\bea
-{\cal L}_{\rm soft} & = & m_i^2 |\Phi_i|^2 + 
(A_{ijk} \l_{ijk} \Phi_i \Phi_j \Phi_k
+ \frac{1}{2} M_a \l^a \l^a + h.c.)
\nonumber \\
& = & m_Y^2|Y|^2+m_{X_1}^2|X_1|^2+m_{X_2}^2|X_2|^2+
(\frac{1}{2}M_{\mu}\lambda^{\mu}\lambda^{\mu}
\nonumber \\
&&+A_1\lambda_1H_1YX_1+A_2\l_2H_2YX_2+...+h.c.),
\eea
where $\Phi_i$ in ${\cal L}_{\rm soft}$ corresponds
to  the scalar component of the corresponding superfield,
$\l^a$ are  gauginos ($\l^{\mu}$ and $M_{\mu}$ are the $\mu$-color
gaugino and its mass, respectively),
and the ellipsis stands for the terms involving only
the MSSM fields. 
In this paper, we will not address the origin of these soft parameters,
but take an approach
to allow generic forms of soft parameters as long as they are
phenomenologically allowed.
In this regard, we  note that string theories with the SUSY
breaking mediated by string moduli show 
enough varieties in the resulting soft parameters\cite{soft1,soft2}.

Let us discuss some global symmetries and the
associated selection rules which will be useful 
for the later discussion of the effective theory below
$\Lambda_{\mu}$. 
In the limit that $W_{\rm tree}$, ${\cal L}_{\rm soft}$,
and  the standard model gauge couplings are all turned
off,  the model is invariant under the $SU(4)$ global
rotation of the four $SU(2)_{\mu}$
doublets $X_{1a}, X_{2a}, Y=(Y_{1a}, Y_{2a})$.
The model includes also several global $U(1)$ symmetries whose charge
assignments are given by
\bea
\label{global}
U(1)_{\rm PQ}: && (Y,X_1,X_2,H_1,H_2,U^c,D^c,E^c, \Lambda_{\mu})
\nonumber \\
&& =(-\frac{1}{2},
-\frac{1}{2},-\frac{1}{2},1,1,  -1,-1,-1, -\frac{1}{2}),
\nonumber \\
U(1)_R: \, && (H_1,H_2, \lambda^a, A_{ijk}, M_a)=(2,2,1, -2,-2),
\nonumber \\
U(1)_{\mu}: \, &&  (Y, X_1,X_2)=(1,-1,-1),
\eea
where the superfields that do not appear in this charge assignment
are understood to have vanishing charge.
Note that $U(1)_{\rm PQ}$ is explicitly broken by
the strong $SU(2)_{\mu}$ anomaly as indicated by
that the holomorphic scale
$\Lambda_{\mu}=M_{GUT}\exp (-\frac{2\pi^2}{g_{\mu}^2(M_{GUT})}+i
\frac{\theta_{\mu}}{4})$
carries nonzero $U(1)_{PQ}$ charge. As a result,
its spontaneous breaking at scales below $\Lambda_{\mu}\sim 1$ TeV
does not lead to any phenomenologically harmful axion.
$U(1)_R$ is free from the $SU(2)_{\mu}$ anomaly,
however broken by the gaugino masses ($M_a$) and $A$-parameters
($A_{ijk}$) carrying $-2$ units of $U(1)_R$ charge.
Finally $U(1)_{\mu}$ corresponds to the $\mu$-baryon number
which is exactly conserved within our framework.

\section{effective theory below $\Lambda_{\mu}$}

In the limit that $m_{\rm soft}\ll \Lambda_{\mu}$ and $\langle H_{1,2}\rangle
\ll \Lambda_{\mu}$, 
light degrees of freedom 
at scales below $\Lambda_{\mu}$ correspond to
$SU(2)_{\mu}$-invariant composite  superfields describing  
$SU(2)_{\mu}$ $D$-flat directions\cite{seiberg}.
In our case, the light composite fields are given by
\bea
Z_{AB}= 
\left(\begin{array}{cccc}
0 & T & Z_{11} & Z_{12} 
\\
-T & 0 & Z_{21} & Z_{22}
\\
-Z_{11} &  -Z_{21} & 0 & S 
\\
-Z_{12} & -Z_{22} & -S & 0
\end{array}
\right)
\eea
obeying the constraint\cite{seiberg}:
\bea
\label{constraint}
{\rm Pf}(Z) = \frac{1}{2} \epsilon^{ABCD} Z_{AB} Z_{CD}
= \epsilon^{\alpha\beta}Z_{1\alpha}Z_{2\beta}-ST
=\hL^2,
\eea
where 
\bea
&& S\sim
\frac{1}{\L_{\mu}}\epsilon^{ab}Y_{1a}Y_{2b},
\quad
T\sim
\frac{1}{\L_{\mu}}\epsilon^{ab}X_{1a}X_{2b},
\nonumber \\
&&
Z_{1\alpha}\sim \frac{1}{\L_{\mu}}\epsilon^{ab}X_{1a}Y_{\alpha b},
\quad
Z_{2\alpha}\sim \frac{1}{\L_{\mu}}\epsilon^{ab}X_{2a}Y_{\alpha b}.
\eea
Here $a,b$ and $\alpha,\beta$  are $SU(2)_{\mu}$ and  $SU(2)_L$ 
doublet indices, respectively.
For the composite fields normalized to have canonical kinetic terms,
the supersymmetric naive dimensional
argument (NDA)\cite{nda1,nda2} leads to
\bea
\hL\ap \L_{\mu}/4\pi.
\eea

The low energy effective action of the  composite
fields $Z_{AB}$ can be expanded in powers of $1/\L_{\mu}$,
more precisely in powers of $H_{1,2}/\L_{\mu}$ and/or  of 
$m_{\rm soft}/\L_{\mu}$, where each term in the expansion
is consistent with the
symmetries and selection rules discussed in the previous section.
The NDA rule\cite{nda1,nda2} then provides an order of magnitude estimate
of the expansion coefficients at
energy scales around $\L_{\mu}$ at which the $SU(2)_{\mu}$ gauge coupling
saturates the bound $g_{\mu} \lesssim 4\pi$.
Let us normalize all superfields to have the canonical
kinetic terms.
Then applying the NDA rule together with the symmetries
and selection rules of the underlying superpotential, 
we find the following form of the effective superpotential
\bea
W_{\rm eff}  = 
X(Z_1Z_2-ST-\hL^2)+a_1 \hL(\l_1 H_1 Z_1 + \l_2 H_2 Z_2)
+W_{\rm MSSM},
\eea
where $W_{\rm MSSM}$ stands for the Yukawa terms involving
only the MSSM superfields, $a_1$ is a {\it nonperturbative} parameter
of order unity, and the $SU(2)_L$ gauge indices are omitted.
Here the Lagrange multiplier superfield $X$ is introduced to
implement the constraint (\ref{constraint}). 
Note that $X$ is not a dynamical field
and so does not appear in the K\"ahler potential.
There may  be additional 
terms in $W_{\rm eff}$ which are higher order in $1/\L_{\mu}$,
but the NDA rule suggests that the effects of such higher order
terms are suppressed by more powers of $\langle H_{1,2}\rangle/\L_{\mu}$.
As will be argued in the subsequent discussions,
$m_{\rm soft}$ and the Higgs VEVs are all comparable
to $\hL\ap \L_{\mu}/4\pi$ in our framework,
and then the $1/\L_{\mu}$ expansion whose coefficients obey
the NDA rule becomes essentially an expansion in powers of $1/4\pi$.
Though not a terribly good approximation, 
we expect that this expansion is reasonably good
and thus the leading order results are not significantly
modified by higher order corrections.

In the $\mu$-color model, there are four doublet  VEVs 
participating in the electroweak symmetry breaking:
\bea
\label{Zmass}
\la H_1 \ra^2 + \la H_2 \ra^2 + \la Z_1 \ra^2 + \la Z_2 \ra^2 
= (178~ {\rm GeV})^2.
\eea
If any of $S$ and $T$ developes a nonzero VEV,
$U(1)_{\mu}$ will be spontaneously broken,
leading to a  potentially dangerous Goldstone boson.
To avoid this problem, we assume 
$\langle S\rangle=\langle T\rangle=0$ which can be easily
achieved by choosing appropriate values of $m_S^2$ and $m_T^2$.
Then the constraint (\ref{constraint})  
gives $\langle Z_1Z_2\rangle=\hL^2$,
and so $\langle Z_1\rangle^2+\langle Z_2\rangle^2\gtr 2\hL^2$.
Furthermore, one would require  $\langle H_2\rangle$ not
significantly smaller than 100 GeV
in order to  avoid a too large top quark Yukawa coupling.
Combining these, one finds
$\hL\lesssim 110$ GeV
where the upper limit is saturated when $\langle Z_1\rangle
\ap \langle Z_2\rangle\ap \hL$.
In most cases, it is phenomenologically  desirable
to have $\hL$ close to its upper limit, and then we have
\bea
\L_{\mu}=4\pi \hL\sim 1 \, {\rm TeV}.
\eea

Soft SUSY breaking terms of the composite fields $Z_{AB}$
can be similarly expanded in powers
of $m_{\rm soft}/\L_{\mu}$
(and also of  $H_{1,2}/\L_{\mu}$)
where $m_{\rm soft}$
denote the soft parameters of the $\mu$-colored elementary fields
renormalized at the NDA scale. 
At the leading order, we find\footnote{
The soft SUSY breaking scalar potential includes 
also the additional $A$-term: $AX(Z_1Z_2-ST-\hL^2)$, but 
this can be eliminated by the redefinition of the
$F$-component of the Lagrange multiplier:
$F_X\rightarrow F_X+AX$.}
\bea
-{\cal L}^{\rm eff}_{\rm soft}&&=m_S^2|S|^2+m_T^2|T|^2+m_{Z_1}^2|Z_1|^2
+m_{Z_2}^2|Z_2|^2
\nonumber \\
&+&(\hat{A}_1\l_1\hL  H_1Z_1+\hat{A}_2\l_2 \hL H_2Z_2+
\hat{A}_3\hL^2 X+h.c.),
\eea
where
\bea
\label{match}
m_S^2&=& a_2(m_{X_1}^2+m_{X_2}^2)+a_3|M_{\mu}|^2,
\nonumber \\
m_T^2&=& 2a_2 m_Y^2+ a_3|M_{\mu}|^2,
\nonumber \\
m_{Z_1}^2&=& a_2(m_{X_1}^2+m_Y^2)+a_3|M_{\mu}|^2,
\nonumber \\
m_{Z_2}^2&=& a_2(m_{X_2}^2+m_Y^2)+a_3 |M_{\mu}|^2,
\nonumber \\
\hat{A}_1&=&a_1A_1+a_4M_{\mu}, 
\nonumber \\
\hat{A}_2&=&a_1A_2+a_4M_{\mu},
\nonumber \\
\hat{A}_3&=& a_5M_{\mu}.
\eea
Here the nonperturbative parameters
$a_i$ ($i=2,3,4,5$) are again of order unity
when the soft parameters of the $\mu$-colored elementary fields 
are renormalized at the NDA scale $\Lambda_{\mu}$
at which $g_{\mu}(\Lambda_{\mu})\sim 4\pi$.

When it is runned from the messenger scale $M_{\rm m}$
of SUSY breaking
to $\L_{\mu}$,
the $\mu$-color gaugino mass 
is {\it enhanced} by the nonperturbative factor  
$\sim 16\pi^2$:
$$
M_{\mu}(\L_{\mu})\sim
\frac{g^2_{\mu}(\L_{\mu}) M_{\mu}(M_{\rm m})}{g^2_{\mu}(M_{\rm m})}
\sim (4\pi)^2 M_{\mu}(M_{\rm m}).
$$
Furthermore if the soft SUSY breaking at $\L_{\mu}$ is dominated
by $M_{\mu}$,
the renormalization group evolution
makes the  other soft parameters of the $\mu$-colored fields
at $\L_{\mu}$, i.e. 
$m_{X_{1,2}}^2$, $m_Y^2$ and $A_{1,2}$,
to be comparable to $M_{\mu}(\L_{\mu})$ also.
Thus if $M_{\mu}$ were comparable to the soft parameters of the
MSSM fields at $M_{\rm m}$, there will arise a $16\pi^2$-hierarchy between
the MSSM soft parameters and the soft parameters of the $\mu$-colored fields
at the NDA scale $\L_{\mu}$, and thus the same hierarchy
between the MSSM soft parameters and the soft parameters of
the composite fields  $Z_{AB}=\{S,T,Z_1,Z_2\}$. 
In order to provide a consistent framework,
the soft parameters of both $Z_{AB}$ and the MSSM  fields
at the electroweak scale
are required to be comparable to $\frac{\L_{\mu}}{4\pi}$.
This means  that at $M_{\rm m}$ the $\mu$-color gaugino mass must be
smaller than the MSSM soft parameters by the factor of $\frac{1}{16\pi^2}$: 
\bea
M_{\mu}(M_{\rm m})\lesssim \frac{1}{16\pi^2}m_{\rm soft}(M_{\rm m}).
\eea
In gauge-mediated SUSY breaking models\cite{dine},
such a small $\mu$-color gaugino mass can  be achieved
if the messenger particles are $SU(2)_{\mu}$-singlets.
In gravity-mediated case, e.g. string effective supergravity models
in which SUSY breaking is mediated by string moduli,
$M_{\mu}(M_{\rm m})$ is small
if the $\mu$-color gauge kinetic function
does {\it not} depend on the messenger moduli 
at string tree level\cite{soft1,soft2}.

\section{higgs phenomenology}
The key difference between the $\mu$-color model and the MSSM
is in the Higgs sector.
To see this,
let us consider the neutral Higgs sector of the model in more detail.
For notational simplicity, in this section let
$Z_{1,2}$ and $H_{1,2}$ denote the neutral components of the 
corresponding composite and elementary Higgs doublets.
Due to the exact $\mu$-baryon symmetry ($U(1)_{\mu}$),
one can always  adjust the parameters of the model,
e.g. $m_S^2$ and $m_T^2$,
to have $\langle S\rangle=\langle T\rangle=0$.
We then have five neutral complex scalar field fluctuations
($\delta\Phi=\Phi-\langle\Phi\rangle$)
with masses of order the 
electroweak scale: 
the two  composite singlet Higgs fluctuations  $\delta S$ and $\delta T$, 
the two elementary doublet Higgs fluctuations
$\delta H_1$ and $\delta H_2$, and 
finally one linear combination
of the composite Higgs doublet fluctuations 
$\delta Z_1$ and $\delta Z_2$ obeying the constraint
$$
\langle Z_1\rangle \delta Z_1+\langle Z_2\rangle \delta Z_2=0.
$$
In particular, we have three physical scalar and two pseudo-scalar
particles arising from the neutral components
of the doublet Higgs fluctuations.

To study the electroweak symmetry breaking and the Higgs mass spectrum,
let us consider the scalar potential of the Higgs doublets
while setting $S$ and $T$ to their vanishing VEVs.
We first have the $F$-term potential 
arising  from the superpotential:
\bea
V_F&=& |XZ_2+\l_1\hL H_1|^2+ |XZ_1+\l_2\hL H_2|^2 \\\nn
&&+|\l_1 \hL^2 Z_1|^2 + |\l_2 \hL^2 Z_2|^2
-(F_X( Z_1Z_2-\hL^2)+h.c.).
\eea
and also the contribution from soft SUSY breaking:
\bea
V_{\rm soft}&=& (\hat{A}_1\l_1\hL H_1Z_1+\hat{A}_2\l_2\hL H_2Z_2+
\hat{A}_3\hL^2 X+h.c.)
\\ \nn
&&+ m_{H_1}^2 |H_1|^2 + m_{H_2}^2 |H_2|^2 
+ m_{Z_1}^2 |Z_1|^2 + m_{Z_2}^2 |Z_2|^2.
\eea
Then the quation of motion for the auxiliary field $X$ yields
\bea
X&=&-\frac{(\l_1 H_1 Z_2^*  + \l_2 H_2 Z_1^*) \hL+
\hat{A}_3^*\hL^{*2}}{ |Z_1|^2+|Z_2|^2},
\eea
leading to 
\bea
V_F+V_{\rm soft}&=&|\l_1\hL Z_1|^2 + |\l_2\hL Z_2|^2
+ |\l_1 \hL H_1|^2 + |\l_2 \hL H_2|^2 \\\nn
&&-\frac{|(\l_1 H_1 Z_2^* + \l_2 H_2 Z_1^*)\hL
+\hat{A}_3^*\hL^{*2}|^2}{|Z_1|^2+|Z_2|^2} 
-(F_X(Z_1Z_2-\hL^2)+h.c.) \\ \nn
&& +(\hat{A}_1\l_1\hL H_1Z_1+\hat{A}_2\l_2\hL H_2Z_2+
+h.c.)
\\ \nn
&&+ m_{H_1}^2 |H_1|^2 + m_{H_2}^2 |H_2|^2 
+ m_{Z_1}^2 |Z_1|^2 + m_{Z_2}^2 |Z_2|^2.
\eea
There is also the $D$-term potential
\bea
V_D&=&\frac{1}{8}(g^2+g^{\prime 2})(|H_1|^2-|H_2|^2-|Z_1|^2+|Z_2|^2)^2.
\eea
Putting these together,
\bea
V&=&V_F+V_D+V_{\rm soft},
\eea
we see  that the Higgs potential takes a form  very different
from that of the MSSM or of other generalizations
of the MSSM.

Since the Higgs potential takes  so different form,
the soft parameter ranges for  successful electroweak 
symmetry breaking can be different also.
Some soft parameter ranges which would not lead
to the correct  electroweak symmetry breaking within the MSSM,
e.g.  {\it positive} $m_{H_1}^2$ and $m_{H_2}^2$ at the electroweak scale,
can successfully  generate the symmetry breaking in the $\mu$-color
framework, while some others which would work in
the MSSM do not work within the $\mu$-color framework.
To see this more explicitly, let us consider the case that
all Higgs doublet VEVs can be chosen to be real.
Then the vacuum stability condition includes
\bea
\left\langle \frac{\partial^2 V}{\partial {\rm Re}(H_{2})^2}
\right\rangle \geq 0,
\eea
which corresponds to 
\bea
2|\l_2 \L|^2 - \frac{2|\l_2 \L|^2 Z_1^2}{Z_1^2+Z_2^2} + 2 m_{H_2}^2
+ \frac{g^2+g^{\prime 2}}{2} ( 3 H_2^2 - H_1^2 + Z_1^2 - Z_2^2 ) \geq 0,
\eea
where all Higgs fields mean their VEVs which are assumed to be real.
Combining this with Eqs. (\ref{constraint}) and (\ref{Zmass})  
which imply ($m_Z=$ the $Z$-boson mass)
\bea
\label{con2}
2 |\hL|^2 \lesssim Z_1^2 + Z_2^2 \lesssim  4 m_Z^2,
\eea
one easily finds  
(with $g^2+g^{\prime 2}\ap 0.5$)
\bea
- m_{H_2}^2 
&\lesssim & 
3 m_Z^2 +|\hL|^2(|\l_2|^2-1-\frac{|\l_2|^2 Z_1^2}{2 m_Z^2})
- \frac{1}{2} Z_2^2
\\ \nn
&\lesssim& 
3 m_Z^2 + (2|\l_2|^2-1) |\hL|^2 -(\frac{|\l_2 \hL|}{m_Z})|\hL|^2.
\eea
For $|\l_2|\lesssim
1.5$, the above limit gives
\bea
m_{H_2}^2
\gtrsim -(174 ~{\rm GeV})^2,
\eea
which is in conflict with the
large portion of the $(\tan\beta, M_{\rm m})$ space
in gauge mediated SUSY breaking models\cite{murayama}
where $M_{\rm m}$ is the messenger scale of SUSY breaking.
This shows that the $\mu$ color model can be incompatible with
certain soft parameter ranges which would be
fine with the conventional $\mu$ term in the MSSM.

Since the Higgs potential of the $\mu$-color model is 
too complicate to get
analytic vacuum solutions for generic parameter values,
here  we  consider two cases one of which
allows an analytic solution, while the other requires
numerical analysis.
The first case is when the parameters renormalized at
the electroweak scale are all real and obey
\bea
\label{para}
&& \l_1\ap\l_2,
\quad \hat{A}_1\ap\hat{A}_2,
\quad m^2_{Z_1}\ap m^2_{Z_2},
\\ \nn
&&m^2_{H_1}\ap m^2_{H_2}\ap \l_1\hL\hat{A_1}>0.
\eea
In this case, it is  straightforward to find that
the Higgs potential has a (local) minimum at
\bea
\la H_1 \ra \ap \la H_2 \ra \ap \la Z_1 \ra \ap \la Z_2 \ra \ap \hL,
\eea
where all parameters are assumed to be real.
The neutral components of the four Higgs doublets,
$H_{1,2}$ and $Z_{1,2}$, constrained as $Z_1Z_2=\hL^2$ contain
three physical scalar and two pesudoscalar particles.
After a tedious but still straightforward computation,
we find the scalar mass eigenvalues are given by
\bea
({\rm scalar \, \, mass})^2\ap (m_{H_1}^2, \, \, m_1^2+m_2^2, \, \, 
m_1^2-m_2^2),
\eea
where
\bea
m_1^2 &=& 2\l_1^2\hL^2+m_{Z_1}^2+(g^2+g^{\prime 2}) \hL^2 \\ \nn
m_2^4 &=& 2\l_1^4\hL^4 +2m_{Z_1}^2\l_1^2\hL^2+
2\l_1^2 (g^2+g^{\prime 2}) \hL^2
+m_{Z_1}^4
\\ \nn
&-&4\l_1^2\hL^2 m_{H_1}^2
-2m_{Z_1}^2 m_{H_1}^2+(g^2+g^{\prime 2})^2 \hL^4,
\eea
and also the pseudoscalar mass eigenvalues
\bea
({\rm pseudoscalar \, \, mass})^2\ap (m_{H_1}^2, \, \, 
2m_{H_1}^2+2\l_1^2\hL^2).
\eea
The $\mu$ color confining scale $\hL \approx 90$ GeV
is fixed by Eq.(\ref{Zmass}), and the Higgs  spectrums are distributed
in  hundred GeV range if the soft parameters 
are also in few hundred GeV range.
The lightest Higgs mass
can be large enough to satisfy the current experimental
lower bound, particularly when the one-loop corrections
involving the large top Yukawa coupling are taken into account.

Different types of VEVs and spectrums are obtained by alleviating 
the relations among the parameters given in (\ref{para}).
Note that the conditions in (\ref{para}),
especially $m_{H_1}^2\ap m_{H_2}^2$ at the electroweak scale,
are difficult to be achieved
in the popular  minimal supergravity model or gauge mediated models,
though possible in string theory models with
moduli-mediated SUSY breaking\cite{soft1,soft2}.
As another example of the successful Higgs phenomenology,
we consider the parameter values at the electroweak scale:
\bea
&& \l_1 \ap 0.5, \quad \l_2 \ap 0.9,
\quad \hat{A_1} \ap \hat{A_2} \ap 70 ~{\rm GeV}, 
\\ \nn
&&
m_{Z_1}^2 \ap m_{Z_2}^2 \ap m_{H_1}^2 \ap (270 ~{\rm GeV})^2,
\quad
m_{H_2}^2 \ap -(30 ~{\rm GeV})^2.
\eea
We then find the Higgs VEVs given by
$$
\la H_1 \ra \ap 7 ~{\rm GeV}, \quad
\la H_2 \ra \ap 120 ~{\rm GeV}, 
$$
\bea
\la Z_1 \ra \ap \la Z_2 \ra \ap \hL \ap 90 ~{\rm GeV},
\eea
and also the masses of the three scalar and two pseudoscalar
neutral Higgs particles
\bea
&&{\rm scalar~ mass}=(90 , \, \, 270 , \, \,  390)~{\rm GeV} 
\nn \\
&&{\rm pseudoscalar~ mass}=(92, \, \, 270)~{\rm GeV}.
\eea
If we include the loop corrections involving the top Yukawa coupling,
the scalar mass can be increased by $ 10 \sim 30 ~{\rm GeV}$ 
depending on the top-stop mass ratio.


\section{conclusion}

In this paper, we introduced a new confining force,
the $\mu$-color, at TeV scale
to replace the $\mu$ parameter in the MSSM superpotential.
Below the $\mu$-color scale,
the model predict composite
Higgs doublets and singlets
whose mass spectrum has been analyzed for certain parameter range.
The $\mu$ color model has very distinctive electroweak
symmetry breaking mechanism
which differs entirely from the conventional
radiatively generated one.
Electroweak symmetry is broken by the 
$\mu$ color dynamics together with soft SUSY breaking terms.
The soft parameter ranges for  successful electroweak 
symmetry breaking can be  quite different from the MSSM
and other generalizations of the MSSM.
Some soft parameter ranges which would not lead
to the correct  electroweak symmetry breaking within the MSSM,
e.g.  {\it positive} $m_{H_1}^2$ and $m_{H_2}^2$ at the electroweak scale,
can successfully  generate the symmetry breaking in the $\mu$-color
framework, while some others which would work in
the MSSM do not work within the $\mu$-color framework.

It would be fair to finally summarize the potentially unattractive
features of the $\mu$-color model which have been noticed
in sections III and IV.
First, we lose the unification of gauge couplings at single scale
due to the  extra $\mu$-colored matter multiplets carrying
$SU(2)_L\times U(1)_Y$ charges.
However this may not be so serious 
in view of that many  string theory scenarios
imply that generically gauge couplings at the string or unification scale
can  take different values.
Second,  in order to implement the electroweak symmetry breaking 
without fine tuning,
it is required that the $\mu$-color gaugino mass at
the SUSY breaking messenger scale $M_{\rm m}$
is smaller than the MSSM soft parameters by the factor of 
$1/16\pi^2$.
Such a small $\mu$-color gaugino mass can  be easily  achieved  within 
gauge-mediated and/or gravity-mediated SUSY breaking models.
In particular,
string effective supergravity models
would give a small $\mu$-color gaugino mass
if the $\mu$-color gauge kinetic function
does {\it not} depend on the messenger moduli 
at string tree level\cite{soft1,soft2}.
Finally the model does
not provide a rationale for $\mu\sim m_{\rm soft}$
since these two mass scales have different dynamical origin.
Even with these features, it appears to be 
worthwhile to study the phenomenological aspects
of the $\mu$-color model in view of its very rich
phenomenologies at  TeV scale.

\section*{Acknowledgments}
We thanks S. Y. Choi and H. B. Kim for helpful
discussions.

\end{document}